\documentclass[aps,pra,10pt,onecolumn,superscriptaddress]{revtex4}
\usepackage{amsmath}
\usepackage{amssymb}
\usepackage{bbm}
\usepackage{graphicx}
\usepackage{framed}
\usepackage{color}
\usepackage{hyperref}
\usepackage{epstopdf}
\usepackage{dsfont}
\usepackage{amssymb}
\usepackage{amsmath}
\usepackage{amsthm}
\usepackage{wrapfig}
\usepackage{relsize}
\usepackage{bm}
\usepackage{textcomp}

 \newtheorem{definition}{Definition}

\mathchardef\minus="002D

\def\<{\langle}\def\>{\rangle}
 \def\ket#1{| #1 \rangle} 
\def\bra#1{\langle #1 |}

\def\v#1{\boldsymbol{\mathrm #1}}

\def\L2{{\mathcal L}_2}

\def\d#1 {\mathop{\!\! \mathrm{d}#1}\,}
\def\df#1#2 {\!\!\frac{\mathop{\mathrm{d}#1}}{#2}\,}

\def\bvec#1{\mathbf{#1}}
\def\bk{\bvec k}
\def\bh{\bvec h}

\def\bn{\bvec n}

\def\bx{\bvec x}

\def\bn{\bvec n}

\def\d{\operatorname{d}}

\def\set#1{\mathsf{#1}}
\def\setH{\set H}
\def\nl#1{\mathcal D^{(#1)}}

\def\Tromb{\set G} 
\def\Ky{\set X} 
\def\Ball{\set{U}}

\def\Sing{\set F} 
\def\GL{\mathbb{GL}}

\begin{document}

\title{Quantum Walks, Weyl equation and the Lorentz group}

\author{Alessandro \surname{Bisio}} \email[]{alessandro.bisio@unipv.it} \affiliation{QUIT group, Dipartimento di
  Fisica, Universit\`{a} degli Studi di Pavia, via Bassi 6, 27100 Pavia, Italy}
\affiliation{Istituto Nazionale di Fisica Nucleare, Gruppo IV, via Bassi 6, 27100 Pavia, Italy}

\author{Giacomo Mauro \surname{D'Ariano}} \email[]{dariano@unipv.it}
\affiliation{QUIT group, Dipartimento di Fisica, Universit\`{a} degli
   Studi di Pavia, via Bassi 6, 27100
  Pavia, Italy}
\affiliation{Istituto Nazionale di
  Fisica  Nucleare, Gruppo IV, via Bassi 6, 27100
  Pavia, Italy}

\author{Paolo \surname{Perinotti}}
\email[]{paolo.perinotti@unipv.it} 
\affiliation{QUIT group, Dipartimento di Fisica, Universit\`{a} degli
   Studi di Pavia, via Bassi 6, 27100
  Pavia, Italy}
\affiliation{Istituto Nazionale di
  Fisica  Nucleare, Gruppo IV, via Bassi 6, 27100
  Pavia, Italy}

% \subtitle{
% \footnote{Work presented at the conference \emph{Quantum
%       Theory: from Problems to Advances}, held on ?????? at at
%     Linnaeus University, Växjö University, Sweden.}}

% \author{Alessandro Bisio \and Giacomo Mauro D'Ariano \and Paolo Perinotti}

% \institute{A. Bisio \and Giacomo Mauro D'Ariano \and Paolo Perinotti \at
% {\em QUIT Group}, Dipartimento di Fisica  and INFN sezione di Pavia, via Bassi 6, 27100 Pavia, Italy\\
%               \email{dariano@unipv.it}}

% % The correct dates will be entered by the editor
\begin{abstract} 
  Quantum cellular automata and quantum walks provide a framework for
  the foundations of quantum field theory, since the equations of
  motion of free relativistic quantum fields can be derived as the
  small wave-vector limit of quantum automata and walks starting from
  very general principles. The intrinsic discreteness of this
  framework is reconciled with the continuous Lorentz symmetry by
  reformulating the notion of inertial reference frame in terms of the
  constants of motion of the quantum walk dynamics. In particular,
  among the symmetries of the quantum walk which recovers the Weyl
  equation---the so called Weyl walk---one finds a non linear
  realisation of the Poincar\'e group, which recovers the usual linear
  representation in the small wave-vector limit.

In this paper we characterise the full symmetry group of the Weyl walk
which is shown to be a non linear realization of a group which is the
semidirect product of the Poincar\'e group and the group of dilations.

% \keywords{Quantum walk\and Doubly Special Relativity}
\end{abstract}

\maketitle

\section{Introduction}

The conjecture, originally advanced by Feynman
\cite{feynman1982simulating}, that the laws of physics can be
ultimately modelled by finite algorithms is a very inspirational
proposal \cite{lloyd2006programming}.  There are many reasons why this
might prove to be the case and, thus, for adopting this conjecture as
a standpoint for a research program.  The primary reason is stated by
Feynman himself: ``It always bothers me that according to the laws as
we understand them today, it takes a computing machine an infinite
number of logical operations to figure out what goes on in no matter
how tiny a region of space and no matter how tiny a region of time". A
similar concern is that in an arbitrarily small region of a continuous
space-time it is in principle possible to store an infinite amount of
bits of information. The only alternative to this situation is that
the dynamics of systems in a finite region of space-time is perfectly
computed by a finite algorithm running on a finite memory.
Furthermore, the idea that the dynamical laws could be reconstructed
within a (quantum) computational framework appears as a natural
continuation of the research on quantum foundations from the
information perspective (see
e.g. Refs. \cite{hardy2001quantum,fuchs2002quantum,quit-derivation,dakic2009quantum}
and for a comprehensive historical overview see Refs.\cite{Khrennikov2012,Khrennikov2015,DAriano20150244}).

As long as we accept that the best microscopic theory at our disposal
is quantum theory, the most natural computational model for the
description of physical laws is a {\em quantum cellular automaton}
\cite{feynman1982simulating,schumacher2004reversible,gross2012index}. The
approach to the foundations of quantum field theory based on quantum
cellular automata was explored for various decades
\cite{bialynicki1994weyl,meyer1996quantum,0305-4470-34-35-323,Yepez:2006p4406}
and it is gathering increasing interest 
\cite{arrighi2014dirac,2016PhRvA..94a2335A,1367-2630-16-9-093007,PhysRevA.94.012335}.
Nevertheless,
the idea that a discrete quantum computer can exactly compute the
evolution of elementary physical systems is seemingly at clash with
continuous symmetries \cite{snyder1947quantized}.

In recent years, free relativistic field equations were derived
starting from the requirements of homogeneity, locality, linearity and
isotropy
\cite{darianopla,PhysRevA.90.062106,Bisio2015244,bisio2014quantum}. The
free quantum field theory (Weyl, Dirac, and Maxwell) is achieved by
restricting to evolutions that are linear in the field--i.e. a
quantum walk--in the limit of small wave-vectors, namely for states so
delocalised that the discrete underlying structure cannot be
resolved. It is remarkable that Lorentz-invariant equations can be
derived without imposing the relativity principle, and not even
mechanical notions. However, the Lorentz symmetry has no direct
interpretation in the above framework, where the geometry of
space-time is not assumed a priori. The achievement of Weyl, Dirac and
Maxwell's equations is a clear indication that an alleged conflict
between discrete dynamics and continuous symmetries was drawn based
only on naive intuition.

In Ref.~\cite{PhysRevA.94.042120} the notion of inertial reference
frame has been formulated in terms of representation of the dynamics
parameterised by the values of the constants of motion. Such notion is
suitable to the study of dynamical symmetries, without the need of
resorting to a space-time background. In this way the Galileo
principle of relativity is formulated by identifying the notion of
change of inertial frame with the change of representation that leaves
the eigenvalue equation of the quantum walk invariant. In the same
Ref.~\cite{PhysRevA.94.042120} it has been shown that such changes of
representations for the Weyl quantum walk encompass a non-linear
realization of the Poincar\'e group. This result, besides embodying a
microscopic model of Doubly Special Relativity (DSR)
\cite{amelino2001planck,PhysRevLett.88.190403,PhysRevD.84.084010},
represents a proof of principle of the coexistence of a discrete
quantum dynamics with the symmetries of classical space-time.

In this paper we review and extend the results of
Ref. \cite{PhysRevA.94.042120} classifying the full symmetry group of
the Weyl quantum walk, which is a semidirect product of the group of
diffeomorphic dilations of the null mass shell by the Poincar\'e
group.

\section{Weyl quantum walk}\label{s:weyl}

A quantum cellular automaton gives the evolution of a denumerable set
of cells, each one corresponding to a quantum system.
We consider the case in which each quantum system is described by
the algebra generated by a set of field operators. Following the
definition of Ref.\cite{schumacher2004reversible}, a quantum
cellular automaton is an automorphism of the
quasi-local algebra. The restriction to non interacting dynamics
corresponds to consider algebra automorphism that are linear in the
field operators (i.e. each field operator is mapped to a linear
combination of field operators). In the same way the dynamics of a
free field is specified by its single particle sector,
a linear quantum cellular automaton is specified by a
quantum walk describing the evolution of a single particle.
A quantum walk \cite{doi:10.1080/00107151031000110776,ambainis2001one}
on a discrete lattice $\Gamma$ of sites $\bx \in \Gamma $ is given by
a unitary operator $A \in \mathcal{L}(\mathcal{H})$ where
$ \mathcal{H} := \ell^2(\Gamma) \otimes \mathbb{C}^{\mathsf{s}} $
where $\ell^2(\Gamma)$ is the space of square summable functions on
$\Gamma$ and $\mathbb{C}^{\mathsf{s}} $ corresponds to some internal
degree of freedom.  If $\ket{\bx}$, $\ket{i}$ are orthonormal basis
for $\ell^2(\Gamma)$ and $\mathbb{C}^{\mathsf{s}} $ respectively, a
(pure)state in $\mathcal{H}$ is a vector
$\ket{\psi} = \sum_{\bx\in \Gamma , i \in {\mathsf{s}} }\psi(\bx,i)
\ket{\bx} \ket{i}$ where
$\sum_{\bx\in \Gamma ,i \in {\mathsf{s}} }|\psi(\bx,i)|^2 =1$.  The
quantum walk $A$ is usually assumed to be local, i.e., for any $\bx$,
we have that $\bra{\bx} \bra{i}A \ket{\bx'} \ket{i'} \neq 0$ only if
$\bx' $ belongs to a finite \emph{neighboring set}\footnote{ For
  example, if $\Gamma$ is the one dimensional lattice which we
  identify with the set of integers $\mathbb{Z}$, we may require
  $|x-x'| \geq n \Rightarrow \bra{x} \bra{i}A \ket{x'} \ket{i'} = 0$
  for some $n\geq 1$. More synthetically we can say that the unitary
  matrix $A$ is block-sparse.}.

As it shown in Ref. \cite{PhysRevA.90.062106} (which we refer to for a
complete discussion),
in the three-dimensional case with minimal dimension ${\mathsf{s}}=2$ 
 the assumptions of \emph{locality},
\emph{homegeneity}, and \emph{isotropy}
single out only one lattice, the body centered cubic one, and four
admissible quantum
walks (modulo a local change of basis) $A^{(\pm)}, B^{(\pm)}$. 
These quantum walks are given by the following unitary 
operators
\begin{align}
  \label{eq:weylqwposition}
  \begin{aligned}
  A^{(\pm)} &= \sum_{\bvec{h} \in S} T_{\bvec{h}} \otimes
  A^{(\pm)}_{\bvec{h}} \\ 
B^{(\pm)} &= \sum_{\bvec{h} \in S} T_{\bvec{h}} \otimes
  B^{(\pm)}_{\bvec{h}}    \quad  B^{(\pm)}_{\bvec{h}}  = (A^{(\pm)}_{\bvec{h}} )^T
  \end{aligned}
\end{align}
where  $S$ is a set of generators of the BCC lattice 
$S:= \{ \pm \bvec{h}_1, \pm \bvec{h}_2, \pm \bvec{h}_3, \pm \bvec{h}_3 \} $
with
\begin{align}
  \begin{aligned}
    &\bh_1=\frac 1{\sqrt3}
    \begin{pmatrix}
      1\\
      1\\
      1
    \end{pmatrix},\ 
    \bh_2=\frac 1{\sqrt3}
    \begin{pmatrix}
      1\\
      -1\\
      -1
    \end{pmatrix},
    &\bh_3=\frac 1{\sqrt3}
    \begin{pmatrix}
      -1\\
      1\\
      -1
    \end{pmatrix},\ 
    \bh_4=\frac 1{\sqrt3}
    \begin{pmatrix}
      -1\\
      -1\\
      1
    \end{pmatrix},
   \end{aligned}
  \label{eq:generators}
\end{align}
$T_{\bvec{h}}$ are the translation operators
$ T_{\bvec{h}}\ket{\bx} = \ket{\bx-\bvec{h}}$, 
and the matrices $  A^{(\pm)}_{\bvec{h}}$ are defined as follows:
\begin{align}
 & \begin{aligned}
A^{(\pm)}_{\bh_1}&=
  \begin{pmatrix}
    \zeta^*&0\\
    \zeta^*&0
  \end{pmatrix},
&A^{(\pm)}_{\bh_2}&=
  \begin{pmatrix}
    0&\zeta^*\\
    0&\zeta^*
  \end{pmatrix},
&A^{(\pm)}_{\bh_3}&=
  \begin{pmatrix}
    0&-\zeta^*\\
    0&\zeta^*
  \end{pmatrix},
&A^{(\pm)}_{\bh_4}&=
  \begin{pmatrix}
    \zeta^*&0\\
    -\zeta^*&0
  \end{pmatrix},
\\ 
A^{(\pm)}_{-\bh_1}&=
  \begin{pmatrix}
    0&-\zeta\\
    0&\zeta
  \end{pmatrix},
 & A^{(\pm)}_{-\bh_2}&=
  \begin{pmatrix}
    \zeta&0\\
    -\zeta&0
  \end{pmatrix},
& A^{(\pm)}_{-\bh_3}&=
  \begin{pmatrix}
    \zeta&0\\
    \zeta&0
  \end{pmatrix},  
&A^{(\pm)}_{-\bh_4}&=
  \begin{pmatrix}
    0&\zeta\\
    0&\zeta
  \end{pmatrix}   
\end{aligned}
  \nonumber\\
&\zeta=\frac{1\pm i}4 \, .
\end{align}
From Eq.~\eqref{eq:weylqwposition}
one immediately sees that the quantum walk commutes with the lattice
translations generated by the vectors $\bvec{h}_i$,
i.e. $[A^{\pm},T_{\bvec{h}_i}\otimes I]=[B^{\pm},T_{\bvec{h}_i}\otimes
I]=0$.
It is therefore convenient to consider the Fourier transform basis
\begin{align}
  \label{eq:fourier}
  \begin{aligned}
    &|\bk\>=\frac1{\sqrt{|B|}}\sum_{\bx\in
  \Gamma}e^{-i\bk\cdot\bx}|\bx\>,
\qquad|\bx\>=\frac1{\sqrt{|B|}}\int_Bd\bk e^{i\bk\cdot\bx}|\bk\>,\\
&\bk=\sum_{j=1}^3k_j\tilde{\bh}_j, \qquad
\tilde{\bh}_j\cdot\bh_l=\delta_{jl} .
  \end{aligned}
\end{align}
where $B$ is the first Brillouin zone of the BCC lattice (see Fig.~\ref{f:brill}).
In the Fourier basis the quantum walks of Eq.~\eqref{eq:weylqwposition}
becomes
\begin{align}
  \label{eq:weylautomaton}
  \begin{split}
&A^{(\pm)}= \int_B d\bk |\bk\>\<\bk|\otimes A^{(\pm)}_\bk,  \qquad
   A_{\v{k}} = I \lambda^{(\pm)}(\v{k}) -i \v{n}^{(\pm)}(\v{k}) \cdot \v{\sigma}^{(\pm)} \\
 &\lambda^{(\pm)} (\v{k}) :=  c_x c_y c_z \mp   s_x s_y s_z \qquad
 \v{n}^{(\pm)} (\v{k})=
 \begin{pmatrix}
   n_x^{(\pm)}     \\
   n_y^{(\pm)}     \\
   n_z^{(\pm)}     
  \end{pmatrix}:=
  \begin{pmatrix}
   s_x c_y c_z   \pm    c_x s_y s_z     \\
   c_x s_y c_z   \mp        s_x c_y s_z     \\
    c_x c_y s_z    \pm   s_x s_y c_z     
  \end{pmatrix}\\
&c_i = \cos  \left( \frac{k_i}{\sqrt{3}} \right) \quad
s_i = \sin \left( \frac{k_i}{\sqrt{3}} \right) \qquad
\v{\sigma}^{(\pm)} := 
(
  \sigma_x ,
\mp\sigma_y ,
\sigma_z 
)^T \,.
  \end{split}
\end{align}
It is possible to show that the matrices $A^{(\pm)}$ can be written as 
\begin{equation}
A^\pm_\bk=e^{-i \frac{k_x}{\sqrt{3}}\sigma_x}e^{\mp i \frac{k_y}{\sqrt{3}}\sigma_y}e^{-i \frac{k_z}{\sqrt{3}} \sigma_z}.
\end{equation}
from which one can immediately see that, in the limit of small
wave-vector $\bk \to 0$, 
the quantum walk $A^{(+)}$ recovers (up to a rescaling
$\frac{\bk}{\sqrt{3}} \to \bk $) the Weyl equation for right-handed
spinors, i.e. $(i \partial_t - \bvec{k} \cdot \bvec{\sigma}) \psi=0$.
Therefore, in order to lighten the notation, it is useful to make the
rescaling
\begin{align}
  \label{eq:rescaling}
  \frac{\bk}{\sqrt{3}} \to \bk.
\end{align}
We can also verify that, in the limit $\bk \to 0$,
the quantum walk $A^{(-)}$ recovers, up to the change of basis induced
by the conjugation with the $\sigma_y$ matrix, the Weyl equation for left-handed
spinors i.e. $(i\partial_t + \bvec{k}\cdot \bvec{\sigma}) \psi=0$.
For this reason, the quantum walks $A^{(\pm)}, B^{(\pm)}$ are called
\emph{Weyl quantum walks}.
The Weyl equation is also recovered when
$ | \bvec{k} - \bvec{k}_i | \to 0$ 
where $\bvec{k}_1 := \tfrac{\pi}{2}(1,1,1)$,
$\bvec{k}_2 := -\tfrac{\pi}{2}(1,1,1)$,
$\bvec{k}_3 := \pi(1,0,0)$. For $\bk \to \bvec{k}_2$
we have the same chirality as for  $\bk \to \bvec{k}_0 :=0 $
while for $\bk \to \bvec{k}_1, \bvec{k}_3 $
the chirality changes. We have then that a single quantum walk
describes four different kind of massless particles,
two left-handed and two right-handed.
This fact can be interpreted as an instance of the known phenomenon of
fermion doubling  \cite{PhysRevD.16.3031} but with a different discrete framework.
In the following we will use the expression ``small wave-vector''
to denote the neighborhoods of the vectors $\bvec{k}_i $,
$i=0,\dots3$.

\subsection{The map $\bvec{n}(\bk)$}\label{sec:map-bvecnbk}

\begin{figure}[h]
  \begin{center}
    \includegraphics[width=\textwidth]{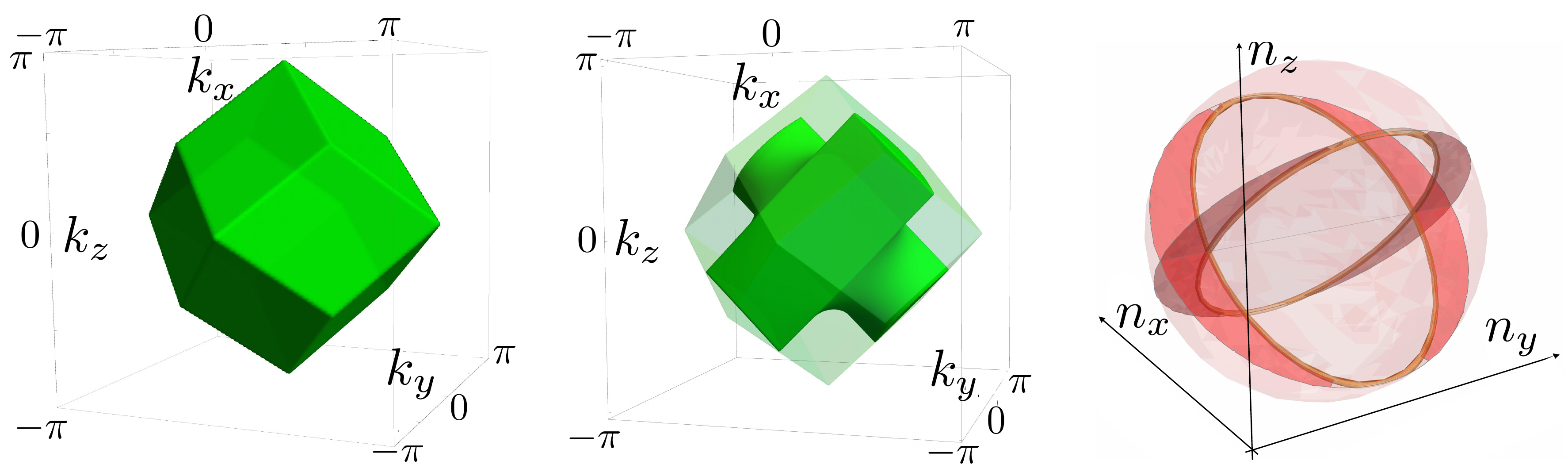}
  \end{center}
\caption{(Colors online) Left: the Brillouin zone for the BCC lattice.
  The region os defined as:
  $B:=\{\bk| -\tfrac{\pi}2\leq
  \bk\cdot\tilde\bh_i\leq\tfrac{\pi}2,\; 1\leq i\leq6\}$, which in
  Cartesian coordinates reads
  $ -\pi\leq k_i\pm k_j\leq\pi,\ i\neq j\in\{x,y,z\}$.
  Middle: The set $B_0 :=  \bvec{n}^{-1}_{0}(\mathsf{H}) $
  embedded in the Brillouin zone.
  Right: the star shaped region $\setH$.
  The set  $\setH$ has been obtained by removing
the set $\Sing'$  (dark red region) from the unit ball.
  \label{f:brill}}
\end{figure}

Before discussing the symmetries and the change of inertial frame for
the Weyl Quantum Walks, we are going to describe some features of
the maps  $\bvec{n}^{(\pm)}(\bk)$ defined in
Eq.~\eqref{eq:weylautomaton}. 
The results we are going to show, will be used for the characterization
of the symmetry transformations of the Weyl Quantum Walks.
For sake of simplicity, we focus on the map $\bvec{n}^{(+)}(\bk)=: \bvec{n}(\bk)$
but the same analysis can be carried out for the map $\bvec{n}^{(-)}$.
Moreover  the map $\bvec{n}(\bk)$ is a smooth analytic map from the Brillouin
zone $B$ to $\mathbb{R}^3$. Its Jacobian $J_{\v{n}}(\v{k})$ is given by
\begin{equation}\label{e:Jacob}
J_{\v{n}}(\v{k}):=\det[\partial_in_j(\v{k})]=\cos(2k_y)\lambda(\v{k}),
\end{equation}
and it vanishes on the set $\Sing :=  \Tromb \cup \Ky$, where
\begin{align}
  \Ky := \{ \v{k} \in B |\ \cos(2 k_y)=0\}, \quad
 \Tromb :=  \{ \v{k} \in B |\ \lambda(\v{k}) = 0  \}.\nonumber
\end{align}
Let us then define the open sets
\begin{align}
  \begin{aligned}
    B'_0 &:=  \{\v{k}\in B |\ \lambda(\v{k}) > 0, \cos(2 k_y) > 0\},\\
 B'_1 &:=  \{\v{k}\in B |\ \lambda(\v{k}) < 0, \cos(2 k_y) > 0\},\\
 B'_2 &:=  \{\v{k}\in B |\ \lambda(\v{k}) > 0, \cos(2 k_y) < 0\},\\
 B'_3 &:=  \{\v{k}\in B |\ \lambda(\v{k}) < 0, \cos(2 k_y) < 0 \}.
  \end{aligned}
\end{align}
and let us denote with $\bvec{n}_{i}(\bk)$ the restriction of
$\bvec{n}(\bk)$
to the set $B'_i $. 
Since $J_{\v{n}}(\v{k}) \neq 0 $ for $\bk \in B'_i$ the map
$\bvec{n}_{i}(\bk)$ defines an analytic diffeomorphism between
$B'_i $ and its image $\bvec{n}_{i}(B'_i )$.
An expression for the inverse map
$\bvec{n}^{-1}_{i} : \mathbb{R}^3 \to  B'_i$
can be obtained exploting the following identities:
\begin{align}
  \label{eq:invfun}
  \begin{aligned}
    & 2(\lambda n_x - n_y n_z ) = \sin 2 k_x \cos 2 k_y, &&  2(\lambda
    n_z - n_y n_x ) = \sin 2 k_z \cos 2 k_y\\
    & 1-2(n^2_x+n^2_y) = \cos 2 k_y \cos 2 k_x, &&  1-2(n^2_z+n^2_y)
    = \cos 2 k_y \cos 2 k_z \\
    & 2( \lambda n_y + n_x n_z) =\sin 2 k_y, && \lambda^2 = 1 - n_x^2
    -n_y^2 - n_z^2.
  \end{aligned}
\end{align}
The ambiguities emerging from the inverse trigonometric functions are
solved by the requirement that  $\bvec{n}^{-1}_{i}(\bvec{n}) \in B'_i$.
One can see that the domain of the inverse function coincides with the
unit ball in $\mathbb{R}^3$ except for the 
image $\bvec{n}(\Sing)$ of the critical points  of $\bvec{n}$.
This set is easily characterized as follows:
\begin{align}
  \begin{aligned}
  &\set {H}':= \Ball \backslash \bvec n({\Sing}),\\
    &\bvec n(\Sing)= \{ \v{m}\in \mathsf{U} |\   m_x = \pm m_z , 2m_x^2 +
    2m_y^2 = 1   \}, \\
  &\mathsf{U} := \{ \v{m}\in\mathbb{R}^3
 |\  \|\bvec{m}\|^2<1  \},
  \end{aligned}
\end{align}
namely the unit ball minus two ellipses (see Fig. \ref{f:brill}).
The map $\bvec{n}_{i} $ then defines an analytic diffeomorphism
between $B'_i$ and $\mathsf{H}'$.
We can easily see that $\mathsf{H}'$ is connected but not simply connected.
For our
purposes we will need to restrict the range of the function
$\bvec{n}$ to a star-shaped (and then simply connected) region.
The largest star-shaped region including $\set{H}'$ is 
\begin{align}
  \begin{aligned}
    &\setH := \Ball\backslash \Sing ',\\
    &\Sing':=\{ \v{m}\in \mathsf{U} |\   m_x = \pm m_z , 2m_x^2 +
    2m_y^2 \geq 1   \}, 
  \end{aligned}
\end{align}
and we also restrict the domain of $\bvec{n}_{i} $ (see Fig. \ref{f:brill}) to the counter image
\begin{align}
B_i :=  \bvec{n}^{-1}_{i}(\mathsf{H}).
\label{eq:invdom}
\end{align}

Let us summarize what we have shown so far.  We have defined four
different sets $B_i$ such that their union is the whole Brillouin zone
$B$ except a null-measure set. We introduced the set $\mathsf{H} $
which is star shaped and differs from the unit ball in $\mathbb{R}^3$
by a null measure set.  For each $i=0,\dots , 3$, the map
$\bvec{n}_i(\bk)$ defines an analytic diffeomorphism between $B_i$ and
$\mathsf{H} $.  We can verify that each of the vectors $\bk_i$, which
were defined at the end of the previous section, belongs to a different
set $B_i$, namely $\bk_i \in B_i$. In the following we will see that we 
can interpret the four regions $B_i$ as the momentum space of four
different massless fermionic particles. 

\section{Change of inertial frame}\label{s:changeframe}
It is now convenient to express the dynamics of the Weyl quantum walk
through its eigenvalue equation
\begin{equation}\label{eq:eigeneq}
A_{\v{k}}\psi(\omega,\v{k})=e^{i\omega}\psi(\omega,\v{k}),
\end{equation}
whose solution set provides an equivalent way to present the walk operator $A$.
In order to lighten the notation we will focus only on the walk $A_{\v{k}}:= A^{(+)}_{\v{k}}$. 
However, the following derivation holds for any of the admissible Weyl quantum walks. 

If we consider the real and imaginary part of $A_{\v k}$ separately, 
Eq.~\eqref{eq:eigeneq} splits into two equations as follows:
\begin{equation}
\left\{\begin{aligned}
&[\cos\omega-\lambda({\v{k}})]\psi(\omega,\v{k})=0,\\
&[\sin\omega I-\bn(\bk)\cdot\v{\sigma}]\psi(\omega,\v k)=0,
\end{aligned}\right.
\label{eq:eigensplit}
\end{equation}
where $\lambda (\v{k})$ and $\bn(\bk)$ were defined in Eq.~\eqref{eq:weylautomaton}.
Notice that the two equations are not independent, as one can easily verify by applying $\sin\omega I+\bn(\bk)\cdot\v{\sigma}$ to the left of the second equation, and then reminding that by unitarity $\lambda(\bk)=1-\|\bn(\bk)\|^2$. The second equation can be easily rewritten in relativistic notation as follows
\begin{align}
\label{eq:hamiltonian2}
n_\mu(k)\sigma^\mu\psi(k) = 0,
\end{align}
where we introduced the four-vectors $k:=(\omega,\bk)$,
$n(k):=(\sin\omega,\v{n}(\bk))$, and we defined
$\sigma:=(I,\v{\sigma})$. The eigenvalues $\omega$ of
Eq.~\eqref{eq:hamiltonian2} then necessarily obey the dispersion relation
\begin{align}
  \label{eq:disprel}
\cos  \omega  = \lambda (\v{k}),
\end{align}
with two branches of eigenvalues, namely
$\omega = \pm \arccos \lambda (\v{k})$.  In the small wave-vector
limit, Eq.~\eqref{eq:hamiltonian2} is approximated by the usual relativistic
dispersion relation $\omega^2 = \|\bk\| ^2$.  Following the analogy with
quantum field theory, we can interpret and the two solutions of
Eq.~\eqref{eq:disprel} as particles for $\omega >0$ and anti-particles
for $\omega<0$. 

Let us now restrict the domain of the function $\bn(\bk)$ to one of
the four region $B_i$ defined in Eq.~\eqref{eq:invdom}.
Since the following considerations won't be affected by the choice of
 $B_i$ we will omit the subscript $i$.
The solutions
of equation \eqref{eq:hamiltonian2} are preserved if we multiply the
left hand side by an arbitrary function $f(k)$ such that
$f(k)\bn(\bk)$ can be inverted as a function on $B_i$.
In particular, we choose an
arbitrary rescaling function $f(k)$ such that $f(k)\bn(\bk)$ maps
$B_i$ to the full $\mathbb R^3$. This is achieved by any rescaling
function $f$ that, besides preserving invertibility of $f(k)\bn(\bk)$
on the regions $B_i$, is singular at the border of the region
$\setH$. In particular, we consider $C^\infty$ functions $f$. The
eigenvalue equation thus becomes

\begin{align}
p^{(f)}_\mu(k) \sigma^{\mu} \psi(k) = 0,\quad p^{(f)}=\mathcal{D}^{(f)}(k):=f(k)n(k) .
\label{eq:poincoveig}
\end{align}

The values $\bk$ and $\omega$ provide a representation of the state space in terms of constants of motion of the quantum walk dynamics. We now define a change of inertial frame as a change of representation that preserve the set of solutions of the eigenvalue equation. We conveniently use the expression of the eigenvalue equation in Eq.~\eqref{eq:poincoveig}.

A change of representation of the dynamics in terms of the constants of motion is given by a function
\begin{align*}
k': k =    \begin{pmatrix}
      \omega\\
     \bk 
    \end{pmatrix}
\mapsto
    k'(k) :=
    \begin{pmatrix}
      \omega'\\
     \bk' 
   \end{pmatrix}.
\end{align*}
We remark that by definition, since $p^{(f)}_\mu(k)=f(k)n_\mu(k)$ and
$n_\mu(k)n^\mu(k)=\det(n_\mu(k)\sigma^\mu)=\sin^2\omega-\|\bn(\bk)\|^2$,
for $\omega= \pm \arccos \lambda (\v{k})$ one has
$p^{(f)}_\mu(k)p^{(f)\mu}(k)=0$. On the other hand, for
$\omega\neq \pm \arccos \lambda (\v{k})$ the eigenvalue equation must
have trivial solution $\psi(k)=0$, and then one has
$p^{(f)}_\mu(k)p^{(f)\mu}(k)\neq0$. Thus, for every invertible map
$k'$ one can define $M(k)\in\GL(2,\mathbb C)$ such that
$M(k)\psi(k)=\alpha(k')\psi(k')$, with $\alpha(k)\in\mathbb{C}$. For
values of $k$ on the mass shell $k=(\omega(\bk),\bk)^T$, this linear transformation
can be expressed in the space
$\ell^2(\Gamma) \otimes \mathbb{C}^{\mathsf{2}}$ as
\begin{align}
T:=\int_{B}\d\!\bk|\bk'(\bk)\>\<\bk|\otimes M(\bk).
\end{align}
Let us  restrict ourselves to
those transformations $k'(k)$ for which there exists an
$M\in\GL(2,\mathbb C)$ independent of $k$ and a rescaling $\alpha(k)$
such that $M\psi(k)=\alpha(k')\psi(k')$.

%Notice that this also implies that the map $k'$ must be continuous.
%  $(k', M, \tilde M  )$ where
%  \begin{align}
%    \label{eq:triple}
%    \begin{aligned}
%&k' :
%k =    \begin{pmatrix}
%      \omega\\
%     \bk 
%    \end{pmatrix}
%\mapsto
%    k'(k) :=
%    \begin{pmatrix}
%      \omega'\\
%     \bk' 
%   \end{pmatrix}\\
%&    M : k \mapsto  M(k) \quad
%  \tilde M : k \mapsto  \tilde M (k) \qquad
%  M(k) , \tilde M(k) \in \GL(2,\mathbb{C})
%  \end{aligned}
%  \end{align}
The above arguments motivate the following definition:
\begin{definition}[Change of inertial reference frame for the Weyl walk]
\label{def:change-inert-frame}
  A change of inertial reference frame for the Weyl walk is a quadruple
  $(k', a, M, \tilde M  )$ where
  \begin{align}
    \label{eq:triple}
    \begin{aligned}
	&k' : k =    
	\begin{pmatrix}
      	\omega\\
     	\bk 
    	\end{pmatrix}
	\mapsto
    	k'(k) :=
    	\begin{pmatrix}
      	\omega'\\
     	\bk' 
   	\end{pmatrix}\\
	&a: B\times[-\pi,\pi]\to[-\pi,\pi]\\
	&  M , \tilde M \in \GL(2,\mathbb{C})
  \end{aligned}
\end{align}
  such that the eigenvalue equation \eqref{eq:poincoveig} is
  preserved, i.e.
  \begin{align}
    \label{eq:changeref}
    p^{(f')}_\mu[k'(k)]\sigma^\mu= \tilde M\,p^{(f)}_\mu(k)\sigma^\mu\, M^{-1},
  \end{align}
  and the eigenvectors are transformed as
  \begin{align}
  \psi'(k')=e^{ia(k)}M\psi(k).
  \label{eq:changeeigvct}
  \end{align}
\end{definition}
Notice that the change of $f$ to $f'$ in Eq. (\ref{eq:changeref})
allows to take $\alpha(k')$ as a phase $e^{i a(k)}$. 
A special case of change of inertial frame is given by the trivial map
$k'=k$ along with the matrices $ M=\tilde{M}=I$. As we will discuss in
the next section, the above subgroup of changes of inertial frame,
that only involves the phases $e^{i a(k)}$, recovers the group of
translations in the relativistic limit. The set of all the admissible changes
of inertial frame forms a group, which is the largest group of symmetries 
of the Weyl walk.

 \begin{figure}[h]
  \begin{center}
    \includegraphics[width=\textwidth]{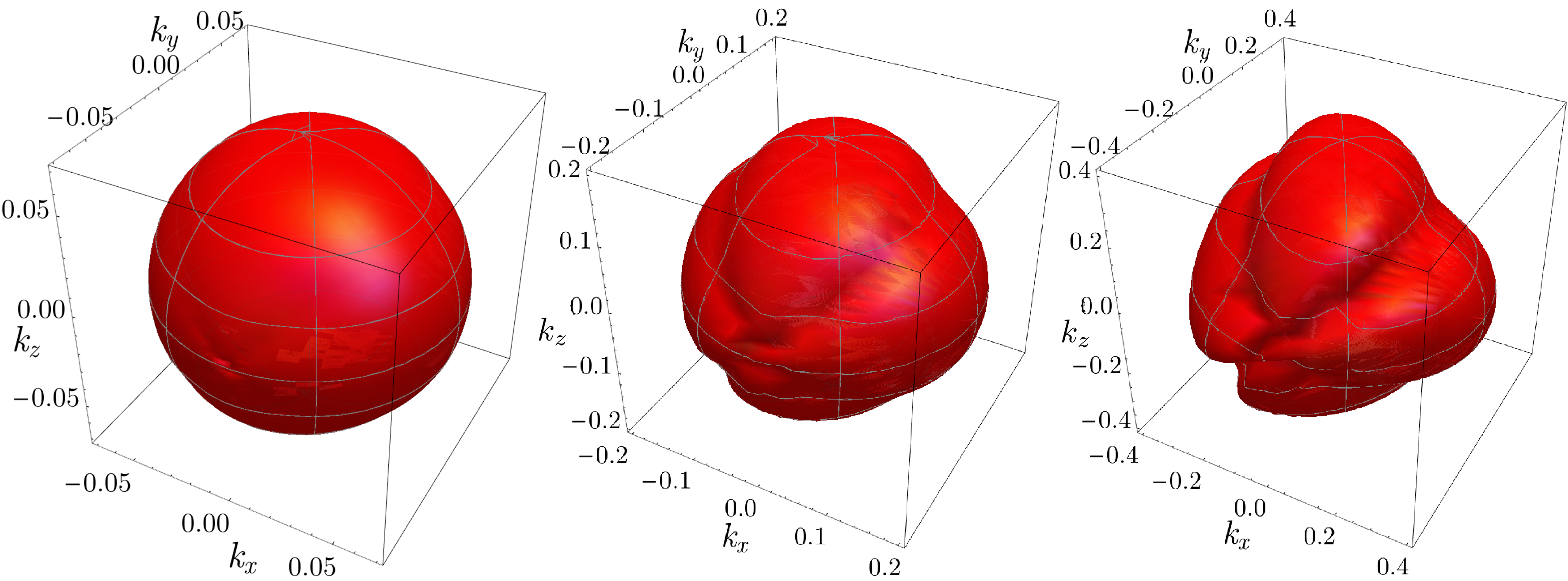}
  \end{center}
  \caption{(Colors online) The red surfaces represent the orbit of a
    wave-vector $\bk = (k_x, 0, 0)$ under the action of the deformed
    rotations
    $\mathcal{R} = {\mathcal{D}^{(f)}}^{-1} \circ R \circ
    \mathcal{D}^{(f)}$
where $f$ is the function defined in Eq.~\eqref{eq:functionf}.
    Left: $k_x= 0.07$.  Middle: $k_x= 0.2$ Right:
    $k_x= 0.4$
    \label{f:pot}}
\end{figure}

In order to classify this group, we now observe that a map acting as
in Eq.~\eqref{eq:triple} transforms the four Pauli matrices linearly
$\sigma^\mu\mapsto L^\mu_\nu\sigma^\nu$, and in turn this implies that
$p^{(f')}_\mu(k')=L_\mu^\nu p^{(f)}_\nu(k)$. Moreover, the set of
invertible linear transformations represented by $L_\mu^\nu$ must
preserve the mass-shell $p^{(f)}_\nu p^{(f)\nu}=0$. By the
Alexandrov-Zeeman theorem  \cite{1964JMP5490Z,Alexandrov:1967fq} this
implies that the transformations $L_\mu^\nu$ must be a representation
of the Lorentz group. Thus, a general change of inertial frame
$(k',a,M,\tilde M)$ for the right-handed Weyl walks must be of the
form
\begin{align}
&k'(k)={\nl g}^{-1}\circ L_\beta\circ \nl f,\nonumber\\
&M=\Lambda_\beta,\quad \tilde M=\tilde\Lambda_\beta,
\label{eq:group}
\end{align}
where $L_\beta$, $\Lambda_\beta $ and $ \tilde\Lambda_\beta $ are the
$(\tfrac12,\tfrac12)$, $(0,\tfrac12)$ and $(\tfrac12,0)$
representations of the Lorentz group, respectively. The only
difference in the case of left-handed Weyl walks is that the
representations $\Lambda_\beta$ and $\tilde\Lambda_\beta$ are
exchanged. Notice that
\begin{equation}
\nl f \circ {\nl g}^{-1}=M_f\circ n\circ n^{-1}\circ M_g^{-1}=
M_f\circ M_g^{-1},
\end{equation}
where
\begin{equation}
M_f(m)=f(n^{-1}(m))m
\end{equation}
one has 
\begin{equation}
\nl f \circ {\nl g}^{-1}(m)=h(m)m,
\end{equation}
and thus
\begin{align}
  \begin{aligned}
&({\nl {g'}}^{-1}\circ L_{\beta'}\circ \nl {f'})\circ({\nl g}^{-1}\circ
L_\beta\circ \nl f)=\\
&({\nl {g'}}^{-1}\circ L_{\beta'}\circ \nl {f'} \circ {\nl g}^{-1}\circ
L_{\beta'}^{-1}) \circ L_{\beta'} \circ 
L_\beta \circ \nl f=\\
&\nl{g''}\circ L_{\beta'\circ\beta}\circ\nl f  \\
&\nl{g''} := {\nl {g'}}^{-1}\circ L_{\beta'}\circ \nl {f'} \circ {\nl g}^{-1}\circ
L_{\beta'}^{-1} 
  \end{aligned}
\end{align}
It is then sufficient to prove that a function $f$ with the desired properties exists, otherwise the group of symmetries of the walk would be trivial.
%We now show that the group of all symmetries of the Weyl walk acts on the Brillouin zone $B$ by a family of nonlinear representations of the Lorentz group, up to arbitrary rescaling functions. 
We have already shown in Section \ref{sec:map-bvecnbk}
that % the Brillouin zone can be split into four regions $B_i$ ($i=0,
% \dots ,3$) such that
the restriction $\bvec{n}_i(\bk)$ of
$\bvec{n}(\bk)$ to $B_i$
define an analytic diffeomorphism between $B_i$  and the manifold
$\mathsf{H} \subset \mathsf{U}$.
%Let us consider the eigenvalue equation~\eqref{eq:hamiltonian2}
%with $\bk \in B_i$, i.e
%\begin{align}
%  \label{eq:restrictedeigeneq}
%  {n}^{\mu}_i(k) \sigma_\mu \psi(k) :=   [\sin \omega I -  \bvec{n}_i(\bk)\cdot  \bvec{\sigma}] \psi(\omega, \bk) =
%  0,
%  \qquad  \bk \in  B_i,
%\end{align}
%where we remind that the restriction to the $B_i$ zones excludes only a null measure set of
%the Brillouin zone.
% Since the following arguments do not depend on $i$, in order to
% lighten the notation,  from now on we will drop the subscript $i$.
Let us consider the solutions of Eq. (\ref{eq:poincoveig}), and define
the function $g(\omega, r \bvec{m} ) := f(\omega, \bvec{n}^{-1}(r \bvec{m}) )$,
where 
%$\overline{m}:=\bvec{n}(\overline{\bk})$,
$g$ is monotonic versus $ r \geq 0$ for every $\bvec m\in\set H$.
We notice that the function $g(\omega, r \bvec{m} )$ is well defined since
$\mathsf{H}$ is star-shaped.
Furthermore, if $g(\omega, r \bvec{m} ) $ diverges on the boundary of $\mathsf{H}$, we have that the map $\mathcal{D}^{(f)}(k)$ defines a diffeomorphism between the set $C_i:=\{ k = (\omega, \bk) |  \bk \in  B_i , \cos \omega = \lambda{\bk} \} $ and the null mass shell $\mathsf{K} := \{ p\in \mathbb{R}^4,  \text{ s.t. }  p^\mu p_\mu = 0 \}$. A possible choice of $f(k) $ which satisfies all the previous requirements is given by
\begin{align}\label{eq:functionf}
\begin{aligned} 
&f(\omega, \bk) :=  f'(\v{n}(\bk)) ,\\
&\tilde{f'}(r,\theta,\phi) := 1+ r \!\!\int_0 ^r \!\!\!\! ds \,\,\left(
  \frac{1}{a(s)} + \frac{1}{b(s,
    \theta, \phi)} \right),\\
&a(r) :=1- r^2 , \quad
 b(r,\theta, \phi) := \cos^22\phi + 
 (\tfrac{1}{2} - r^2 (1- \cos^2\theta \sin^2\phi))^2
  \end{aligned} 
  \end{align}
where we used spherical coordinates $n_x = r \cos\theta \cos\phi$,
$n_y = r \sin\theta $, $n_z = r \cos\theta \sin\phi$ for the argument
in the definition of the function $f':\set H\to\mathbb R$, with the
convention that for $\bn=\bvec0$ one has $\phi=0$.
% , and thus
% $b(\bvec0)=1$, so that $f'(\bvec0)=1$ (for this purpose, it is
% actually sufficient that $b(\bvec0) \neq 0 $).

In order to classify the most general transformation leaving the walk invariant, it is still possible to allow for transformations of the kind
\begin{align}
&k'=\sum_i n_{j(i)}^{-1}n_i,\\
&a=0,\quad M=\tilde M=I,
\end{align}
where the region $B_i$ is mapped to the region $B_{j(i)}$. Notice that this corresponds to a permutation of the four regions $B_i$, which however must fulfil the constraint that $i$ and $j(i)$ must labe lregions corresponding to walks with the same chirality ($\{B_0,B_2\}$ and $\{B_3,B_4\}$). This part of the group thus corresponds to $\mathbb Z_2\times \mathbb Z_2$.

By considering the case $f=g$
in Eq.~\eqref{eq:group}, we have
\begin{align}
  \label{eq:deflor}
  \mathcal{L}_{\beta}:= {\mathcal{D}^{(f)}}^{-1} \circ L_\beta  \circ {\mathcal{D}^{(f)}}
\end{align}
which is a non linear representation of the Lorentz group as the ones
considered within the context of doubly special
relativity \cite{amelino2001planck,amelino2002relativity,magueijo2003generalized}.
It is easy to observe that, if $f'(0)=1$ and $\partial_\mu f' = 0$
where $f(\omega,\bk)=f'(\sin\omega, \bvec{n}(\bk))$ as in
Eq.~\eqref{eq:functionf}, the Jacobian $J_{\mathcal{L}_{\beta}} $ of
$\mathcal{L}_{\beta}$ coincides with $L_\beta$. In the limit of small
wave-vector we have that $\mathcal{L}_{\beta} = L_\beta + O(|\bk|^2)$
that is the non linear Lorentz transformations recover the usual
linear one.  In Fig.~\ref{f:pot} we show the numerical evaluation of
some wave-vector orbits under the subgroup of rotations  of the
nonlinear representation of the Lorentz group. We see how the
distortion effects, which are negligible for small wave-vector, become evident at
larger wave-vectors.

\section{Conclusion}\label{s:concl}

The analysis of the previous section can be in priniciple applied to
any quantum walk dynamics for which we know a complete set of constant
of motion.  In particular we could consider the Dirac quantum walk of
Ref. \cite{PhysRevA.90.062106}, whose eigenvalue equation
is $(p_\mu(\omega, k, m) \gamma^\mu - m I)\psi(\omega, k, m) = 0$
where $\gamma_\mu$ are the Dirac $\gamma$ matrices in the chiral
representation, $m$ is the particle mass and
$p(\omega, k, m):=(\sin\omega, \sqrt{1-m^2} \bvec{n}(\bk) )$.
In this case we may generalize Definition \ref{def:change-inert-frame}
and allow for maps that change the value of $m$. We can then consider
the invariance of the whole family of Dirac quantum walks parametrized by $m$.
One could prove that the symmetry group of the Dirac walks include a
non-linear representation of the De Sitter group $\mathbb{SO}(1,4)$.

Since the frequency (or energy) $\omega$ and the wave-vector (or
momentum) $\bk$
are the constant of motion of the quantum walk dynamics, 
the scenario we discussed so far deals with the changes of reference frame
in the energy-momentum $(\omega ,\bk )$ space.
In particular we saw that the Lorentz group is recovered and
one could wonder how to give a
time-position description of
the deformed relativity framework
that we obtained in energy-momentum space.
It is believed that the nonlinear deformations of the Lorentz group in
momentum space have profound consequences on our notion of space-time.
In particular we may have the emergence of relative locality  \cite{PhysRevD.84.084010},
i.e. the coincidence of events in space-time becomes observer dependent.
This would imply that not only the coordinates on space-time are
observer dependent, as in ordinary special relativity, but also that
different observer may infer different space-time manifolds for the
same dynamics.
Non-commutative space-time and Hopf algebra
symmetries  \cite{lukierski1991q,majid1994bicrossproduct,kowalski2002doubly,kowalski2003non,Bisio20150232}
have been also considered for a time-position space formulation
of deformed relativity.

\begin{acknowledgements}
This publication was made possible through the support of a grant from
the John Templeton Foundation under the project ID\# 60609 {\em
  Quantum Causal Structures}. The opinions expressed in this publication are those of the authors and do not necessarily reflect the views of the John Templeton Foundation.
\end{acknowledgements}

%\bibliographystyle{spphys}
%\bibliographystyle{spmpsci}
%\bibliography{bibliography}

% \begin{thebibliography}{10}
% \providecommand{\url}[1]{{#1}}
% \providecommand{\urlprefix}{URL }
% \expandafter\ifx\csname urlstyle\endcsname\relax
%   \providecommand{\doi}[1]{DOI \discretionary{}{}{}#1}\else
%   \providecommand{\doi}{DOI \discretionary{}{}{}\begingroup
%   \urlstyle{rm}\Url}\fi

% \end{thebibliography}

%
% QUA VA PIAZZATA LA BIBLIOGRAFIA COME LA VUOLE FOOP
%  COPIO E INCOLLO IL BBL
%

% \begin{thebibliography}{10}
% \providecommand{\url}[1]{{#1}}
% \providecommand{\urlprefix}{URL }
% \expandafter\ifx\csname urlstyle\endcsname\relax
%   \providecommand{\doi}[1]{DOI~\discretionary{}{}{}#1}\else
%   \providecommand{\doi}{DOI~\discretionary{}{}{}\begingroup
%   \urlstyle{rm}\Url}\fi

% \end{thebibliography}

% \bibliographystyle{spmpsci}
% \bibliography{bibliography}

\bibliographystyle{apsrev4-1}
\bibliography{bibliography}

\end{document}